\documentclass{elsart}

    \usepackage{latexsym,bm,amsmath,amssymb,amsfonts}
    \usepackage{epsfig,graphics,graphicx}
    \usepackage{makeidx}
    \usepackage{citesort}

    \renewcommand{\theequation}{\thesection.\arabic{equation}}

\long\def\comment#1{ }

\newcommand{\beq}{\begin{eqnarray}}
\newcommand{\eeq}{\end{eqnarray}}
\newcommand{\be}{\vspace{-.4cm}\begin{eqnarray}}
\newcommand{\ee}{\vspace{-.5cm}\end{eqnarray}}

\newcommand{\cal}{\mathcal} %ELS%
\newcommand{\tr}{{\rm tr}}
\newcommand{\Tr}{{\rm Tr}}

\newcommand{\BQ}{\begin{equation}}
\newcommand{\EQ}{\end{equation}}
\newcommand{\BQA}{\begin{eqnarray}}
\newcommand{\EQA}{\end{eqnarray}}

\def\simge{\mathrel{%
   \rlap{\raise 0.511ex \hbox{$>$}}{\lower 0.511ex \hbox{$\sim$}}}}
\def\simle{\mathrel{
   \rlap{\raise 0.511ex \hbox{$<$}}{\lower 0.511ex \hbox{$\sim$}}}}

\begin{document}

\begin{flushright}
~\vspace{-1.25cm}\\
 BNL-NT-05/45

\end{flushright}
\vspace{2.cm}

\begin{frontmatter}

\parbox[]{16.0cm}{ \begin{center}
\title{On the Wess--Zumino term in high energy QCD}

\author{Yoshitaka ~Hatta}

\address{ RIKEN BNL Research Center, Brookhaven National Laboratory,
Upton, NY 11973, USA}

\date{\today}
%\vspace{0.8cm}
\begin{abstract}
Recently, Kovner and Lublinsky proposed new small--$x$ QCD
evolution equations valid when the gluon density inside the target
is low. The key element of their construction
 is the Wess--Zumino term which ensures  the non-commutativity
 of valence charges. In this paper we clarify the origin and significance of
 this term by showing that it can be naturally
  incorporated in the effective theory of Color Glass Condensate.
 We also reexamine the renormalization group description in the high
 density (JIMWLK) regime.
\end{abstract}
\end{center}}

\end{frontmatter}

\section{Introduction}
\setcounter{equation}{0} \label{S_Intro}

Over the past few years, considerable effort has been made
\cite{rare,shoshi,stat,edmond,MSW05,KL1,KL2,KL5,Blaizot,ian,HIMST,dipole,numerical}
towards understanding the high energy  QCD evolution equation
beyond the Balitsky--Kovchegov (BK) equation \cite{B,kov}, or the
Jalilian-Marian--Iancu--McLerran--Weigert--Leonidov--Kovner
(JIMWLK)
 equation \cite{JKLW97,RGE}. Interest in the problem was
sparked by several seemingly unrelated observations
\cite{salam,rare,shoshi,stat,edmond} regarding the insufficiency
of the BK--JIMWLK equation, but they all point to a necessity to
include new types of Feynman diagrams called \emph{Pomeron loop}
diagrams.

\begin{figure}
\begin{center}
\centerline{\epsfig{file=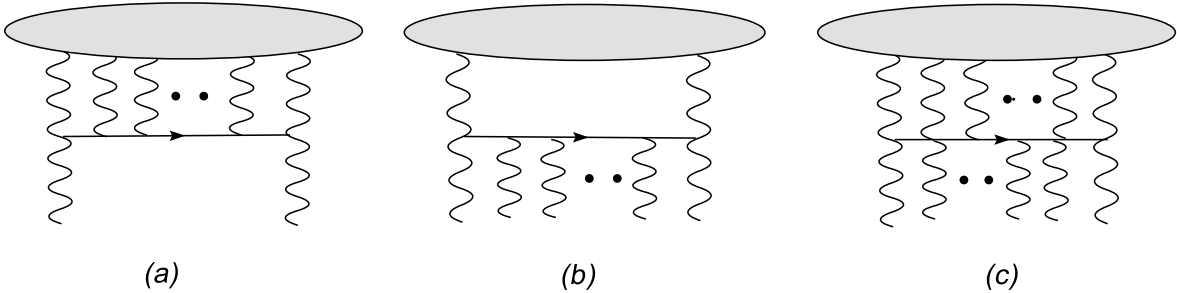,height=4.cm}} \caption{\sl
Quantum evolution of the target (represented by a blob): (a) Gluon
recombination (b) Gluon splitting (c) Both the gluon recombination
and the splitting
 \label{fig1}}
\end{center}
\end{figure}

 Fig.~\ref{fig1}(a) is a typical diagram
summed by the JIMWLK equation. The upper blob represents an
energetic hadron (the ``target") moving in the positive $z$
direction. At very high energy, the target behaves like a
 weakly coupled many body system of small--$x$ gluons commonly dubbed the Color
Glass Condensate (CGC). In one step of quantum evolution towards
smaller $x$, arbitrarily many $t$--channel gluons inside the
target recombine into two gluons. This is the \emph{gluon
saturation} phenomenon \cite{levin,larry,Lipatov,reviews} which
plays an important role for the unitarization of the BFKL Pomeron
\cite{BFKL}.

 Attention is currently focused on
   diagrams with arbitrarily many gluon legs as represented by Fig~1(b).
   Viewed from the target side, these diagrams describe the
   \emph{gluon splitting} or the \emph{gluon number fluctuation}
   inside the target \cite{edmond}.
   They are particularly important
    when the target is dilute, and are responsible for the
    eventual formation of a dense system. On the other hand, the same diagrams,
    when viewed from the projectile
    side, clearly describe the gluon recombination process in the
    projectile. In either interpretation, Fig.~1(b) is the leading
    diagram in the kinematic regime which is complementary to that considered
    in the JIMWLK equation.
 A set of evolution equations including the vertex of
Fig.~\ref{fig1}(b) with four gluon
 legs below the rung has been proposed as the first step beyond the
 JIMWLK equation \cite{edmond,MSW05,dipole}. [See also earlier works \cite{Bartels}.]
  These results were obtained in the frameworks of
    Mueller's dipole model \cite{Mueller} and its alternative
    formulation as a color glass \cite{IM}.
 In this model, the relevant gluon splitting process (or better be called the Pomeron
splitting process in the large--$N_c$ approximation) is
  naturally
 described by the dipole splitting and subsequent gluon emissions.
 Combining the dipole model with the JIWMLK formalism in the
  dipole sector \cite{janik}, one can investigate
  more general problems which involve both the Pomeron splitting and
the Pomeron recombination (Fig.~1(c)), i.e., the Pomeron loops
\cite{Blaizot,Bartels}.

Meanwhile, in a very interesting paper \cite{KL1} Kovner and
Lublinsky have
  derived evolution equations in the dilute regime to \emph{all}
orders in the gluon legs without using the large--$N_c$
 approximation. The kernel of the evolution equation was shown to
 be \emph{dual} to the JIMWLK kernel \cite{ian,KL2,HIMST} essentially reflecting the symmetry
  between Figs. 1(a) and 1(b). However, identification of states (``observables") on which this kernel acts
   does not follow from simple duality considerations.
 The nature of the evolution
  equation in the dilute regime is peculiar; one has to treat color charges as \emph{non-commutative}
operators.\footnote{As shown in \cite{dipole}, this problem
disappears in the dipole model in the large--$N_c$ approximation,
and one can check the consistency with the previous results in
\cite{edmond}.} Kovner and Lublinsky overcame this problem by
introducing a
  \emph{Wess--Zumino term}  and an extra coordinate (``ordering
  variable"), after which the charges can be treated as commutative.
  The emergence of this term at first sight is
   somewhat mysterious, as it is rarely discussed in the context of high energy QCD.
 Still, the fact that the duality and the
  relationship between the wave function approach \cite{KL1} and
 the effective action/Hamiltonian approaches
  \cite{ian,HIMST,Lipatov} become manifest only in this
 ``commutative world"  calls for
 a formulation which explicitly includes the Wess--Zumino term.
  [See, however, \cite{KL5}.]

 In this paper we
fully  clarify the origin and significance of the Wess--Zumino
term in the \emph{high} density (JIMWLK) regime. This may be
unexpected as the issue of non-commutativity seems least important
in this regime. On the contrary, we shall show that the
Wess--Zumino term is quite useful and can be naturally
incorporated in the JIMWLK formalism as a part of the source term.
This observation has led us to  more detailed treatment of  the
\emph{renormalization group} description  than in the literature
\cite{JKLW97,RGE} in the sense that we discuss renormalization of
the source term. It should be emphasized that, although in this
paper we focus our attention to the JIMWLK regime, our approach is
general and therefore applicable  to both the dilute regime
(Fig.~\ref{fig1}(b)) and the Pomeron loop regime
(Fig.~\ref{fig1}(c)) where the importance of the Wess--Zumino term
has been first recognized. Indeed, an essential difference between
 effective theories in these regimes and the current JIMWLK
formalism \cite{JKLW97,RGE} is that one has to introduce the
light--cone time coordinate $x^+$($=$ordering variable) in the
former \cite{HIMST}. By reformulating the latter with
$x^+$--dependent classical charges, we can address the high energy
evolution in different regimes in a single framework.
 In this context the Wess--Zumino term necessarily arises as a consequence of
 gauge invariance.

 In Section 2, after a brief review of the
JIMWLK formalism we propose a new source term which explicitly
 includes a Wess--Zumino term. We show that our source term
correctly reproduces the induced charge upon quantum evolution. In
Section 3, we perform the renormalization group evolution directly
at the level of the JIMWLK functional integral. We conclude with
Section 4.

\section{JIMWLK formalism with the Wess--Zumino term}
\setcounter{equation}{0}

\subsection{The Color Glass Condensate} \label{2.1}
 The color glass condensate (CGC) formalism \cite{larry,JKLW97,RGE,reviews} is an effective theory of gluon
 saturation at a small Bjorken parameter $x=\Lambda^+/P^+$, where
 $P^+$ and $\Lambda^+$ are the light--cone momenta of the right--moving parent
 hadron (the target) and the small--$x$ gluon of interest, respectively. Partons
 with momentum fraction larger than $\Lambda^+$ are described as the
 external source $\rho$, while the gluons with momenta $\Lambda^+$
 are described as the classical field $A^{\mu}$ created by the
 source according to the Yang--Mills equation \begin{align} D_\nu
 F^{\nu \mu}=\delta^{+\mu}\rho(\vec{x}). \label{YM} \end{align}
 We use the notation $\vec{x}\equiv (x^-,x^i)$ where $x^i \ (i=1,2)$ is
 the two--dimensional transverse
  coordinate. In the saturated
 regime (JIMWLK regime), $\rho$ is effectively \emph{static} (i.e.,
 $x^+$--independent) and parametrically of order $\sim
 \cal{O}(1/g)$.
 In the light cone gauge $A^+=0$, the solution to Eq.~(\ref{YM}) is given by
 \begin{align} A^\mu= \delta^{\mu i}B^i, \ \ \ \ \
 B^i=\frac{i}{g}U\partial^i U^{\dagger}, \label{back} \end{align}
 where $U$ is a Wilson line in the $x^-$ direction  \begin{align}
 U^\dagger(\vec{x})={\mbox P}\exp\left(ig\int^{x^-}_{-\infty} dz^-
 \alpha(z^-,x^i) \right). \label{wil} \end{align}
$\alpha$ is the classical solution in the Coulomb gauge
\begin{align} \tilde{A}^\mu \equiv \delta^{\mu+}\alpha, \label{cou}
\end{align} and is related to the Coulomb gauge charge $\tilde
\rho \equiv U^\dagger \rho U$ via the plus component of the
Yang--Mills equation
\begin{align} -\nabla^2_\perp \alpha=\tilde{\rho}. \end{align}
 Observables  ${\cal O}$ (e.g., scattering amplitudes)  are first calculated
 for
 a given background field Eq.~(\ref{back}), ${\cal O}[\rho]$, and
 then averaged over $\rho$ with the \emph{weight function} $W_\tau[\rho]$
 \begin{align} \langle {\cal O} \rangle_\tau=\int D\rho W_\tau[\rho] {\cal
 O}[\rho], \label{observable}
 \end{align} where $\tau$ is the rapidity $\tau=\ln 1/x=\ln P^+/\Lambda^+$.
 $W_\tau$
 satisfies the JIMWLK
 equation \cite{JKLW97,RGE}
 which is a renormalization group
 equation in rapidity \begin{align} \frac{\partial}{\partial
 \tau}W_\tau[\rho]=-H_{\mbox{\scriptsize{JIMWLK}}}\left[\alpha, \frac{\delta}{\delta \rho}\right]
 W_\tau[\rho]. \label{JIM} \end{align}
 The precise form of the JIMWLK Hamiltonian $H_{\mbox{\scriptsize{JIMWLK}}}$ is irrelevant
 here. [See, however, Eq.~(\ref{JIMWLK}).]
 Suffice it to say that it is quadratic in the functional
 derivative $\delta/\delta\rho$ (corresponding to the two gluon legs in
 Fig.~\ref{fig1}(a))
 and all orders in $\alpha$ (corresponding to the merging gluons above the
 rung in Fig.~\ref{fig1}(a))
  in the form of the Wilson line
 Eq.~(\ref{wil}).

 The derivation of Eq.~(\ref{JIM}) proceeds with the following
 steps \cite{JKLW97,RGE}: (i) Start with the QCD functional
 integral in the light--cone gauge
\begin{align} \int D\rho W_\tau[\rho] \int_\tau DA^\mu
\delta(A^+)e^{iS_{\mbox{\scriptsize{YM}}}[A^\mu]+iS_{\mbox{\scriptsize{W}}}[A^\mu,
\rho]}, \label{functional}
\end{align} where the subscript $\tau$ in the $A^\mu$ integral means that the gauge
fields contain only modes with $p^+ < \Lambda^+=e^{-\tau}P^+$.
Modes with $p^+ > \Lambda^+$ have been already integrated out.
$S_{\mbox{\scriptsize{W}}}$ is the \emph{source term} which gives
the right hand side of the Yang--Mills equation Eq.~(\ref{YM}).
(ii) Expand the gauge field around the solution to the Yang--Mills
equation
\begin{align} A^\mu=\delta^{\mu i}B^i+a^\mu+\delta A^\mu,
\label{ex}
\end{align} where $a^\mu$ is the \emph{semihard} field with the
momentum fraction $\Lambda^+ > p^+> b\Lambda^+$ ($b\ll 1$) and
$\delta A^\mu$ is the \emph{soft} field with $p^+ < b\Lambda^+ $.
(iii) Functionally integrate out the semihard field $a^\mu$. (iv)
Show that, in the leading logarithmic approximation (LLA), i.e.,
keeping terms proportional to $\alpha_s \ln 1/b\equiv \alpha_s
\delta \tau$ ($\alpha_s=g^2/4\pi$), the effect of the integration
is absorbed by the renormalization of the weight function
$W_\tau[\rho]\to W_{\tau+\delta \tau}[\rho+\delta \rho]$, where $
\delta \rho$ is the additional charge induced by the semihard
field. In this paper we
 perform this program in the most direct way, including the
  \emph{renormalization of the source term} $S_{\mbox{\scriptsize{W}}}$.

For this purpose, first we must know $S_{\mbox{\scriptsize{W}}}$.
The original definition in \cite{JKLW97} is \begin{align}
S_{\mbox{\scriptsize{W}}}=\frac{i}{N_c}\int d\vec{x} \tr
[\rho(\vec{x})\widetilde{W}(\vec{x})] \label{ori},
\end{align}
 where \begin{align} \widetilde{W}(\vec{x})\equiv {\mbox
 P}\exp\left(ig\int_{-\infty}^{\infty}
 dx^+A^-_a(x^+,\vec{x})T^a \right), \label{w} \end{align} with
 $T^a$  the color matrices in the adjoint representation.
  [Hereafter $W$ and $\widetilde{W}$ denote
  Wilson lines in the fundamental and adjoint representation,
  respectively.]
  Eq.~(\ref{ori}) is invariant under $x^+$--independent gauge
  transformations. Upon quantum evolution,
 one shifts $A^- \to a^-+\delta A^-$ [Note that $A^-=0$ for the classical solution.]
 and expands the exponential. The coefficient of $-\delta A^-$
  defines the \emph{induced charge} $\delta \rho[a^-]$ to be included in the
 classical theory at rapidity
  $\tau+\delta\tau=\ln P^+/b\Lambda^+$. Several different
choices of $S_{\mbox{\scriptsize{W}}}$ can be found in the
literature \cite{RGE,raju}. The non-uniqueness of
$S_{\mbox{\scriptsize{W}}}$  simply means that they are
\emph{effective}  actions. Although they all lead to the
renormalization group equation Eq.~(\ref{JIM}), it is not clear
whether intermediate calculations go hand-in-hand with the
corresponding perturbative QCD calculation.
 Moreover, when we consider the \emph{all} order effect of $A^-$
 in the dilute regime, the choice of $S_{\mbox{\scriptsize{W}}}$ is crucial to obtain
  correct evolution equations.

Below we construct another $S_{\mbox{\scriptsize{W}}}$ that is
most suited for our purpose. This is \emph{not} in the form of an
effective action, but contains its own dynamics due to the
presence of the Wess--Zumino term. For the sake of simplicity and
clarity, throughout this paper we consider the color SU(2) gauge
group.\footnote{The form of the Wess--Zumino term is a bit simpler
for SU(2) than for SU(3). The latter can be found in the
literature. See, for example,  Ref.~\cite{diakonov}.} We describe
the target hadron as a bunch of energetic ``valence partons"
having longitudinal momenta $p^+>e^{-\tau}\Lambda^+$. They carry
various representations of SU(2)
($J=\frac{1}{2},1,\frac{3}{2},\cdots$) at various
 transverse coordinates $x^i$, and are also distributed in a narrow
 strip
 in the $x^-$ direction with the width $|\Delta x^-|\sim 1/\Lambda^+$ (Fig.~2).
 By a deliberate choice of the light--cone
 prescription, one can restrict the support of the charges to the
 region
 $x^->0$ \cite{RGE}.

\begin{figure}
\begin{center}
\centerline{\epsfig{file=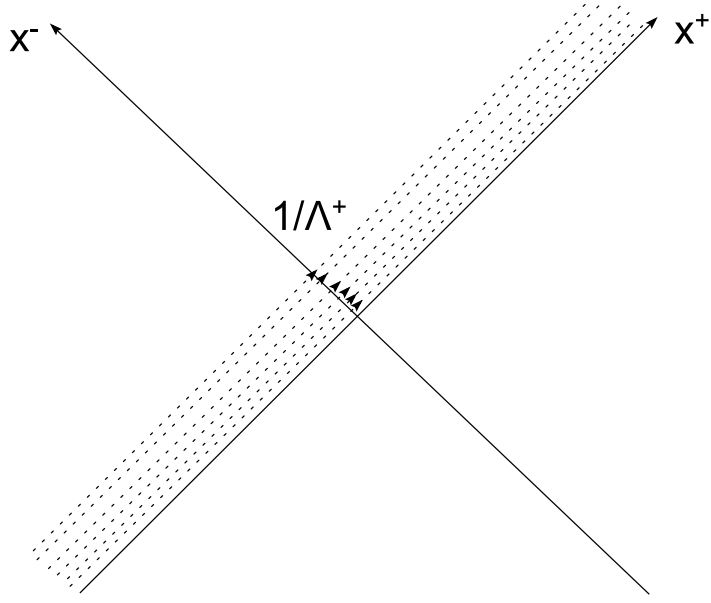,height=8.cm}} \caption{\sl
Eikonal propagation of the valence charges. The charges have
support only at positive $x^-$.
 \label{fig2}}
\end{center}
\end{figure}

 Let us focus on a single quark at the coordinate $\vec{x}$ inside the target.
 In the eikonal approximation, the propagation of this quark is described by the
 amplitude
\begin{align} Z=\langle \bar{\psi}_aW^{ab}(\vec{x})\psi_b\rangle,
\label{quark}
\end{align} where $W$ is the Wilson line  Eq.~(\ref{w}) in the fundamental
representation $T^a \to \tau^a/2$ ($\tau^a$ $a=1,2,3$ are the
Pauli matrices) and
 $\psi$ is the Dirac spinor of the quark.
 With the use of a formula for an \emph{open } Wilson
 line $W^{ab}$ derived by Diakonov and Petrov \cite{diakonov},
 Eq.~(\ref{quark}) takes the form
\begin{eqnarray} Z&=&\int
DS_{\pm \infty}W[S_{\infty},S_{-\infty}]
  \int_{S_{-\infty}}^{S_{\infty}}DS(x^+)\int DA^\mu  \nonumber \\&& \qquad \qquad  \times \exp\left(iS_{\mbox{\scriptsize{YM}}}
  + igJ\int
dx^+ \tr
[\tau_3SA^-S^{\dagger}]+iS_{\mbox{\scriptsize{WZ}}}[S(x^+)]
\right) \nonumber
\\ &&=\int
D\rho_{\pm \infty}W[\rho_{\infty},\rho_{-\infty}]
\int_{\rho_{-\infty}}^{\rho_{\infty}}D\rho(x^+) \int DA^\mu
\nonumber \\ && \qquad \qquad \qquad \times
\exp\left(iS_{\mbox{\scriptsize{YM}}}- i\int dx^+ \rho(x^+)
A^-(x^+) +iS_{\mbox{\scriptsize{WZ}}}[\rho(x^+)] \right),
\label{min}
\end{eqnarray}
 where $S(x^+)$ is a SU(2) matrix ($S_{\pm \infty}\equiv S(x^+=\pm \infty)$), $J=1/2$ for the quark fundamental representation.
  $S_{\mbox{\scriptsize{WZ}}}$ is
 the \emph{Wess--Zumino term } (also called a geometric phase, Berry's phase, Polyakov's spin
  factor) \cite{wz,diakonov} \begin{align}
 S_{\mbox{\scriptsize{WZ}}}=iJ\int^{\infty}_{-\infty} dx^+ \tr \, [\tau^3
 S\partial^-S^\dagger], \label{wess}
 \end{align} where $\partial^-=\partial_+=\partial/\partial x^+$.
In the second equality in Eq.~(\ref{min}), we switched to the
$\rho$--representation defined by ($A^-=A^-_a\frac{\tau^a}{2}$)
   \begin{align} \rho^a(x^+)=-\frac{gJ}{2}\tr \ [ \tau^3S\tau^a
S^\dagger]. \label{rho1}\end{align}  The constraint
$\rho^a\rho^a=g^2J^2$ is implicit in the measure $D\rho$. For a
Wilson \emph{loop}, $S_{\mbox{\scriptsize{WZ}}}$ can be  written
as a functional of $\rho$ by introducing yet another coordinate
$u$ $(1\ge u \ge 0)$ (which is the radial coordinate of the disc
spanned by the loop) and extrapolating  $\rho(x^+) \to
\rho(x^+,u)$ such that  $\rho(x^+,u=1)=\rho(x^+)$.
   But for an open Wilson line, this is not possible in general.\footnote{ We
could consider an overall color--singlet target
  and make closed loops by connecting Wilson lines at different transverse coordinates
  at $x^+=\pm \infty$. However, this leads to unnecessary complications
   because in the light--cone gauge in which we are working, there
   is nonzero transverse field $B^i$ at infinity. }
 Nevertheless, we use the notation
 $S_{\mbox{\scriptsize{WZ}}}[\rho]$ with $\rho$ and $S$ related by
 Eq.~(\ref{rho1}). The ``weight function" $W[\rho_{ \infty},\rho_{-\infty}]$ (not to be confused
 with the Wilson line $W$) contains
 Wigner rotation matrices \cite{diakonov} and quark spinors. It is in general complex and depends on $S$ (or
 $\rho$) only at $x^+=\pm \infty$.
Note that the sum of the last two terms in the exponential of
Eq.~(\ref{min})
\begin{align} gJ \tr
[\tau_3SA^-S^{\dagger}]+ iJ\tr \, [\tau^3 S\partial^-S^\dagger]
\label{inv}
\end{align} is invariant under the following gauge transformation
 \begin{align} \tilde{
 A}^-=U^{\dagger} A^- U +\frac{i}{g}U^{\dagger}\partial^-U,
 \qquad \
\tilde{S}=SU \label{tran} \end{align} Under this transformation,
$\rho$ transforms as
\begin{align} \tilde{\rho}=U^{\dagger}\rho U. \label{rhotran}
\end{align}

 In fact, Eq.~(\ref{min}) is a  \emph{path integral formula for a spin} \cite{wz}.
  The exponential factor \begin{align} H(x^+)=-gA^-_a(x^+)\frac{\tau^a}{2}, \end{align} of the fundamental
  Wilson line in Eq.~(\ref{quark}) may be regarded as a Hamiltonian
   for a non-relativistic ``spin" $\tau^a$ immersed in a time--dependent ``magnetic field"
   $A^-_a(x^+)$. The Wess--Zumino term is nothing but the kinetic
   term (the $p\dot{q}$ term in Eq.~(\ref{kine})) which arises when
   one goes from the Hamiltonian to the Lagrangian\footnote{ One can make the correspondence
    explicit by introducing angular coordinates  $\rho \propto (\sin \theta \cos
 \phi, \sin \theta \sin \phi, \cos \theta)$. Under this parametrization, $p \sim \cos \theta$, $q \sim \phi$ and
 $D\rho \sim DpDq$. } \begin{align}
   \mbox{P}\exp\left(-i\int dx^+ H\right) \sim \int DpDq
   \exp \left( i\int dx^+ (p\dot{q}-H)\right). \label{kine}
 \end{align}

 Generalization to an arbitrary set of valence partons (with longitudinal momenta
 $p^+>e^{-\tau}\Lambda^+$) can be done simply
by replacing \begin{align} J \to \int d\vec{x}J(\vec{x}),
\end{align} and endowing  $W_\tau[\rho_{\pm\infty}(\vec{x})]$ with the information about
the distribution of partons and their
 group representations. This can be done along the lines suggested in
 \cite{raju2}.
 In this way we are led to
the following\emph{ source term} ($x=(x^+,\vec{x})$)
\begin{align} S_{\mbox{\scriptsize{W}}}[A^-,\rho]=- \int dx \rho(x) A^-(x)
+S_{\mbox{\scriptsize{WZ}}}[\rho(x)], \label{sou} \end{align} with
$\rho(x)$ defined by Eq.~(\ref{rho1}) with $J\to J(\vec{x})$.
Correspondingly, Eq.~(\ref{min}) is  generalized to
\begin{align} Z =\int D\rho_{\pm
\infty}(\vec{x})W_\tau[\rho_{\infty},\rho_{-\infty}]
\int_{\rho_{-\infty}}^{\rho_{\infty}}D\rho(x^+,\vec{x})\int_\tau
DA^\mu
\exp\left(iS_{\mbox{\scriptsize{YM}}}[A^\mu]+iS_{\mbox{\scriptsize{W}}}[A^-,\rho]\right).
\label{start}
\end{align} Eq.~(\ref{start}) is the starting point of quantum
evolution.  By extending the definition of gauge transformation to
the charge sector, Eq.~(\ref{rhotran}), one can explicitly
maintain the gauge invariance of the theory. Namely,
Eq.~(\ref{start}) can also be written as
\begin{align} Z =\int D\tilde{\rho}_{\pm
\infty}W_\tau[\tilde{\rho}_{\infty},\tilde{\rho}_{-\infty}]
\int_{\tilde{\rho}_{-\infty}}^{\tilde{\rho}_{\infty}}D\tilde{\rho}(x^+)
\int_\tau D\tilde{A}^\mu
\exp\left(iS_{\mbox{\scriptsize{YM}}}[\tilde{A}^\mu]+iS_{\mbox{\scriptsize{W}}}
[\tilde{A}^-,\tilde{\rho}]\right). \label{start2}
\end{align}

\subsection{Induced charge from the source term} \label{2.2}

In order for Eq.~(\ref{sou}) to be an acceptable source term,
first we have to show that it gives the correct induced charge
$\delta\rho[a^-]$ under one step of quantum evolution $A^- =
a^-+\delta A^-$. Namely, $S_{\mbox{\scriptsize{W}}}[a^-+\delta
A^-,\rho]\sim -(\rho+\delta \rho[a^-])\delta A^-$. At first sight
this looks impossible because $S_{\mbox{\scriptsize{W}}}$ is
linear in $A^-$ (so $a^-$ and $\delta A^-$ do not couple).
However, we observe that the corresponding coupling $\bar{\psi}
\gamma^+ t^a \psi A^-_a \sim \rho^a A^-_a$ in the QCD Lagrangian
  is also linear in $A^-$. As with the perturbative QCD calculation,
  we expand the exponential $e^{-i\rho A^-}$ in powers of $ A^-$. To
quadratic order, we get
\begin{align} -\frac{1}{2}\int D\rho(x^+) \int dx^+dy^+
\rho_a(x^+)\rho_b(y^+)e^{iS_{WZ}}(\delta
A^-_a(x^+)+a^-_a(x^+))(\delta A^-_b(y^+)+a^-_b(y^+)), \label{aa}
\end{align} Hereafter we suppress the spacial coordinate $\vec{x}=(x^-,x^i)$.
The  crucial step of our approach is that we rewrite the $a^-
\delta A^-$ coupling in Eq.~(\ref{aa}) as
  \begin{align} -\int D\rho(x^+) \int dx^+dy^+
\rho_a(x^+)\rho_b(y^+)\delta A^-_a(x^+)a^-_b(y^+) e^{iS_{WZ}}
\nonumber \\ =-\int dx^+dy^+ \left(\theta(x^+-y^+)\langle
\hat{\rho}^a\hat{\rho}^b\rangle +\theta (y^+-x^+)\langle
\hat{\rho}^b\hat{\rho}^a\rangle \right)\delta
A^-_a(x^+)a^-_b(y^+), \label{ik}
\end{align} where $\hat{\rho}^a$ ($a=1,2,3$) are non-commutative charge
operators. In the zero dimensional problem,
$\hat{\rho}^a=-g\tau^a/2$. In the three--dimensional case
$\hat{\rho}^a$ are  functions of $\vec{x}$ and satisfy local
commutation relations \begin{align} [\hat{\rho}^a(\vec{x}),
\hat{\rho}^b(\vec{y})]=-ig\epsilon^{abc}\hat{\rho}^c\delta^{(3)}(\vec{x}-\vec{y}).
\label{cmm}
\end{align}
 Eq.~(\ref{ik}) is the `magic' of the Wess--Zumino term which can be mathematically
 justified \cite{wz}. A path integral of
 commutative $\rho$'s endowed with
  a Wess--Zumino term is equal to a matrix element of
  non-commutative
    charges
  $\hat{\rho}$. The ordering of $\hat{\rho}$'s follows from the
  ordering of the $x^+$ coordinate of $\rho(x^+)$ under the path integral.

 By decomposing the product of two $\hat{\rho}$'s into the symmetric and anti-symmetric parts \begin{align}
\hat{\rho}^a\hat{\rho}^b=\frac{1}{2}[\hat{\rho}^a,
\hat{\rho}^b]+\frac{1}{2}\{\hat{\rho}^a, \hat{\rho}^b\},
\end{align} we write the  second line of Eq.~(\ref{ik}) as
 \begin{align} -\frac{1}{2}\int dx^+dy^+
\left(-igf^{abc}[\theta(x^+-y^+)-\theta
(y^+-x^+)]\langle\hat{\rho}^c\rangle+\langle\{\hat{\rho}^a,\hat{\rho}^b\}\rangle
\right)\delta A^-_a(x^+)a^-_b(y^+). \label{ver}
\end{align}
 The symmetric term
$\propto \langle\{\hat{\rho}^a,\hat{\rho}^b\}\rangle$ in
Eq.~(\ref{ver}) vanishes because it is proportional to
\begin{align} \int dy^+ a^-_b(y^+) =a^-_b(p^-=0)=0. \label{zero} \end{align}
 Eq.~(\ref{zero})  is valid  since the semihard field $a^\mu$ is nearly an on--shell
excitation \cite{RGE} having $\Lambda^+
> p^+>b\Lambda^+$ and $p_\perp^2/2b\Lambda^+ > p^- > p_\perp^2/2\Lambda^+$,
where $p_\perp$ is a typical transverse momentum. [In the LLA, the
precise value of $p_\perp$ does not matter.]  Returning to the
path integral representation $\langle \hat{\rho}^c\rangle \to \int
D\rho \,\rho^c e^{iS_{\mbox{\scriptsize{WZ}}}}$, we
re-exponentiate Eq.~(\ref{ver})  and read off (a part of) the
induced charge from the coefficient of $-i\delta A^-$
\begin{align} \delta \rho^{(1)}_a(x^+) \equiv -\frac{gf^{abc}}{2}\rho^c_{-\infty}\int dy^+
 \big(\theta(x^+-y^+)-\theta (y^+-x^+)\big)
 a^-_b(y^+). \label{vert}
 \end{align} The $x^+$ coordinate of $\rho^c$ can be chosen freely, since
  there is only one factor of $\hat{\rho}$ left in Eq~(\ref{ver}). This really does not matter;
    as we shall argue at the beginning of Section 3, $\rho$ is almost static in the JIMWLK
regime. In the above, we have
  set $x^+=-\infty$.
  Eq.~(\ref{vert}) agrees with the literature \cite{JKLW97,RGE}, although the
  derivation here is  very different. As remarked above, our derivation is closely tied to the corresponding
   pQCD calculation. Indeed,
  pQCD, the $a^-\delta A^-$ coupling arises due to second order
  perturbation \begin{align} -\frac{g^2}{2}(\bar{\psi}\gamma^+ t^a  \psi \delta
A^-_a) ( \bar{\psi}\gamma^+ t^b  \psi a^-_b) \sim -\frac{1}{2}
\rho^a\delta A^-_a \rho^b a^-_b,
\end{align}
 We see that the Wess--Zumino term is a convenient trick to replace the dynamical
 fermionic field with classical charges while correctly retaining the
 color commutator which in pQCD was implied by the quantum commutator.

 Next we consider the virtual contribution to the induced charge
 coming from the
 $\sim \rho a^- a^- \delta A^-$
coupling. This time we expand up to the cubic order
\begin{eqnarray}
&&\frac{(-i)^3}{3!}\int D\rho\left( \rho (a^- + \delta A^-)
\right)^3 e^{iS_{WZ}} \nonumber \\ &&\ \ \ \sim \frac{i}{2}\int
D\rho \int dx^+ dy^+dz^+ \rho^a(x^+)\rho^b(y^+)\rho^c(z^+)
\delta A^-_a(x^+)a^-_b(y^+)a^-_c(z^+) e^{iS_{WZ}} \nonumber \\
&&\ \ \ =\frac{i}{2}\int dx^+ dy^+dz^+  \delta
A^-_a(x^+)a^-_b(y^+)a^-_c(z^+)\Bigl( \theta_{xyz}
\langle\hat{\rho}_a\hat{\rho}_b\hat{\rho}_c\rangle +\theta_{xzy}
\langle \hat{\rho}_a\hat{\rho}_c\hat{\rho}_b\rangle \nonumber \\
&&\ \ \ \ \ \ \ \ \ \ \ \ \ \ \ \ \ +
\theta_{yxz}\langle\hat{\rho}_b\hat{\rho}_a\hat{\rho}_c\rangle
+\theta_{yzx}\langle\hat{\rho}_b\hat{\rho}_c\hat{\rho}_a\rangle
+\theta_{zxy}\langle\hat{\rho}_c\hat{\rho}_a\hat{\rho}_b\rangle
+\theta_{zyx}\langle \hat{\rho}_c\hat{\rho}_b\hat{\rho}_a\rangle
\Bigr),
\end{eqnarray}
where \begin{align} \theta_{xyz} \equiv \theta(x^+-y^+)
\theta(y^+-z^+). \end{align} The coefficient of $-i\delta A^-$ is
another contribution to the induced charge $\delta \rho^{(2)}$.
For this part of the induced charge,  all we need in the later
developments is the expectation value $\langle \delta
\rho^{(2)}\rangle$
 where the averaging $\langle
\dots \rangle$ is computed in the Gaussian approximation using the
background field propagator Eq.~(\ref{mod}).  Anticipating this,
we perform the replacement
 \begin{align}  a^-_b(y^+)a^-_c(z^+) \to
 \delta_{bc}iG^{--}(y^+-z^+) \label{paper2}
\end{align} already at this point. Note that this component of the
propagator is diagonal in color indices. This is certainly correct
for the free propagator (which is the case for the BFKL regime).
In the JIMWLK regime, the background field modifies the
propagator. Nevertheless, Eq.~(\ref{paper2}) is valid in practice.
See, Appendix C of the third paper in Ref.~\cite{RGE}.
  Since $\hat{\rho}_b\hat{\rho}_b$ is a group Casimir, it commutes
with $\hat{\rho}_a$. Using this fact and Eq.~(\ref{zero}), we
obtain after some algebra
\begin{align}
-\frac{ig^2N_c}{4}\int dx^+ dy^+dz^+  \delta
A^-_a(x^+)iG^{--}(y^+-z^+)\hat{\rho}_a \bigl(
\theta(y^+-x^+)\theta(x^+-z^+)+\theta(z^+-x^+)\theta(x^+-y^+)\bigr).
\end{align} From this we can read off the (expectation value of the) induced charge
\begin{align} \langle \delta \rho^{(2)}_a\rangle \equiv \frac{g^2N_c}{4}\rho^a_{-\infty} \int dy^+dz^+
iG^{--}(y^+-z^+) \bigl(
\theta(y^+-x^+)\theta(x^+-z^+)+\theta(z^+-x^+)\theta(x^+-y^+)\bigr),
\label{cub}
\end{align}
 in agreement with the literature.

Summarizing, we expanded $e^{-i\rho (a^- + \delta A^-)}$ and
re-exponentiated  the relevant
 (in the sense of the LLA) couplings between $a^-$ and $\delta A^-$.
 Effectively, we have replaced
 %\footnote{Our choice of the $x^+$
 %argument of $\rho$'s in the second line is somewhat arbitrary, as they are
 %not unambiguously determined in our exponentiation scheme.
 %Similar ambiguity arises also in the
 %Yang--Mills part, namely, whether one expands around the static
 %(Eq.~(\ref{back}))
 %or $x^+$--dependent (Eq.~(\ref{YM3})) background fields in Eq.~(\ref{small}). These are not real
 %problems since, as we argue later, $\rho$ is almost static in the JIMWLK regime. }
   \begin{eqnarray}
&&\exp \left( -i\int dx^+ \rho(x^+)\big(a^-(x^+)+ \delta
A^-(x^+)\big) \right) \nonumber \\ &&\to  \exp \left(-i\int dx^+
\rho_{-\infty} a^-  -i\int dx^+ \big( \rho(x^+)+
\delta\rho^{(1)}+\langle \delta \rho^{(2)}\rangle \big) \delta A^-
+\frac{i}{2}\int a^-\Pi a^- \right),\nonumber  \\  \label{replace}
\end{eqnarray} where we again chose $\rho(x^+=-\infty)$ in the linear term in
$a^-$ (see the remark below Eq.~(\ref{vert})).
 The last term in Eq.~(\ref{replace})  is
 a correction to the background field propagator (c.f., Eq.~(\ref{mod})) coming from the
 $a^-a^-$ term in Eq.~(\ref{aa})
 \begin{align} \frac{i}{2}\int a^-\Pi a^-\equiv \frac{igf^{abc}}{4}\int dx^+dy^+
\big(\theta(x^+-y^+)-\theta (y^+-x^+)\big)\rho^c_{-\infty}
a^-_a(x^+)a^-_b(y^+). \label{sign}
\end{align}

\section{Quantum evolution and renormalization group at high density}
\setcounter{equation}{0}

\subsection{Recovering the CGC picture}

In this section, we revisit the renormalization group evolution in
the high density regime \cite{JKLW97,RGE} with the aim of showing
 the renormalization of the source term. We start with
Eq.~(\ref{start}) in the light--cone gauge $A^+=0$ and expand
around the solution to the classical equations of motion which are
obtained by varying the field $A^{i,-}$
\begin{align} D_\mu F^{\mu +}=\rho(x^+), \ \ \ D_{\mu } F^{\mu
i}=0. \label{YM2}
\end{align}
 Eq.~(\ref{YM2}) was solved in \cite{HIMST}, with the solution again in the same
 form as Eq.~(\ref{back})\footnote{ In \cite{HIMST}, the $x^+$--dependence
 of $\rho$ was assumed to be of the form
   Eq.~(\ref{saddle}). However, the solution
  Eq.~(\ref{YM3}) is more general and is valid for arbitrary
  $x^+$--dependence of $\rho(x^+)$.}
 \begin{align} A^\mu(x^+)=\delta^{\mu i}B^i(x^+), \ \ \ \
  B^i=\frac{i}{g}U\partial^i U^{\dagger}(x^+). \label{YM3} \end{align}
  Crucial simplification in the
JIMWLK regime is that we can neglect the $x^+$--dependence in
Eqs.~(\ref{YM2}) and (\ref{YM3}). This is because the important
paths in the path integral $\int D\rho(x^+)$ are those with weak
$x^+$ dependence. To illustrate this, consider the solution to the
saddle point equation
\begin{align}
D^-\rho(x^+)=0. \label{san}
\end{align}
This can be obtained by performing infinitesimal rotation
\begin{align} S \to Se^{i\omega}, \qquad \omega=\omega^a\frac{\tau^a}{2}, \end{align}
 in Eq.~(\ref{wess}) and Eq.~(\ref{rho1}). Under this
 rotation, $\rho$ and $S_{\mbox{\scriptsize{WZ}}}$ change by
 \begin{align} \delta \rho^a=\epsilon^{abc}\omega^b \rho^c,
 \qquad \delta S_{\mbox{\scriptsize{WZ}}} =-\frac{1}{g}\int dx^+ \rho^a \partial^-
 \omega^a. \end{align} Eq.~(\ref{san}) immediately follows by imposing $-\int \delta
 \rho^a A^-_a + \delta S_{\mbox{\scriptsize{WZ}}}=0$.
The solution to Eq.~(\ref{san}) is
\begin{align} \rho^a(x^+)=\widetilde{W}_{ab}(x^+)\rho_{-\infty}^b=
\bigl(W(x^+)\rho_{-\infty} W^\dagger(x^+)\bigr)^a, \label{saddle}
\end{align}
 where \begin{align} W(x^+)=\mbox{P}\exp\left( ig\int^{x^+}_{-\infty}
dz^+ A^-(z^+)\right). \end{align} In Appendix \ref{A2}, we derive
an interesting property of the saddle point solution
Eq.~(\ref{saddle}). Since $A^-$ field is weak in the JIMWLK
regime, at the saddle point
\begin{align} \rho(x^+) \approx \rho_{-\infty}. \label{app} \end{align}
 At high density where $\rho
\sim {\cal O}(1/g) \gg 1$,  the path integral receives important
contributions from the
 neighborhood of the saddle point. Therefore, it is legitimate to
  approximate $\rho(x^+)\approx \rho_{-\infty}$ when solving  Eq.~(\ref{YM2}).
 Moreover, \begin{align} W_\tau[\rho_\infty, \rho_{-\infty}]
 \approx  W_\tau[\rho_{-\infty}, \rho_{-\infty}]\equiv W_\tau[\rho_{-\infty}]. \label{wreal}
 \end{align} This
 is a very important approximation which allows us to recover the CGC picture
 from our $x^+$--dependent formulation. Namely,
 as a result of Eq.~(\ref{wreal}), $W_\tau$ becomes
 real and positive, therefore it literally serves as a weight
 function.
   In the high density regime,  at \emph{fixed} rapidity, the above argument about the saddle point is
    generally correct and leads to a
    static,  classical theory characterized by the averaging Eq.~(\ref{observable}). However,
    the saddle point solution and the
corresponding
    approximation
    Eq.~(\ref{app})  have only a limited sense when we consider quantum
    evolution,
     namely, when the semihard field $a^\mu$ is introduced.
 Indeed, it was essential to keep track of the $x^+$--dependence of
$\rho(x^+)$ and maintain the full (not saddle point) path integral
$D\rho(x^+)$ with the Wess--Zumino term in order to correctly
derive the induced charge $\delta \rho^{(1,2)}[a^-]$.
 Therefore, we use the approximation $\rho(x^+)\approx
\rho_{-\infty}$ only when it is safe to do so. Our criterion of
`safe' is that the semihard field $a^-$ is not involved. For
example, $W_\tau[\rho_{\infty},\rho_{-\infty}]$ is safe but the
term $-\rho(x^+)a^-(x^+)$ in $S_{\mbox{\scriptsize{W}}}$ is not:
$\rho(x^+)a^-(x^+) \neq \rho_{-\infty}a^-(x^+)$. In this way we
can maximally exploit the approximately static nature of the
charges in the JIMWLK regime without tampering the precise $x^+$
structure of the source term
$S_{\mbox{\scriptsize{W}}}[\rho(x^+)]$.

%A simple mnemonic is that $\rho(x^+) \delta A^- \approx
%\rho_{-\infty}\delta A^-$, but $\rho(x^+) a^- \neq \rho_{-\infty}
%a^-$ (c.f., Eq.~(\ref{zero})).

\subsection{Renormalization group}

 With these  caveats in mind, we expand around the static solution
 $B^i[\rho_{-\infty}]$ \begin{eqnarray} &&Z=\int D\rho_{-\infty}
 W_\tau [\rho_{-\infty}]\int_{\rho_{-\infty}}^{\approx \rho_{-\infty}}D\rho(x^+)
 \int_\tau D\delta A^{i,-}
\nonumber \\
 && \qquad \qquad \times \exp\left(iS_{\mbox{\scriptsize{YM}}}[B^i+\delta A^i, \delta A^-]- i\int
dx^+ \rho \delta A^- +iS_{\mbox{\scriptsize{WZ}}}[\rho(x^+)]
\right). \label{fin}
\end{eqnarray} In the above, the field $\delta A^{i,-}$ contains
modes with $p^+<\Lambda^+$.
 In order to derive the effective theory at rapidity $\tau +
 \delta \tau$, we
  decompose the soft field $\delta A^{i,-}\to a^{i,-}+\delta
  A^{i,-}$ and functionally integrate over $a^{i,-}$  \begin{eqnarray} &&Z=\int D\rho_{-\infty}
 W_\tau [\rho_{-\infty}]\int_{\rho_{-\infty}}^{\approx \rho_{-\infty}}D\rho(x^+)
 \int_{\tau+\delta \tau}D\delta A^{i,-} \int_{\tau}^{\tau+\delta \tau} Da^{i,-}
\nonumber \\
 &\times& \exp\left(iS_{\mbox{\scriptsize{YM}}}[B^i+a^i+\delta A^i, a^-+\delta A^-]- i\int
dx^+ \rho (a^-+\delta A^-) +iS_{\mbox{\scriptsize{WZ}}}[\rho(x^+)]
\right) \label{22}
\end{eqnarray}

In the source term we perform the replacement Eq.~(\ref{replace}).
The Yang--Mills action gives other contributions \cite{JKLW97,RGE}
to the induced charge
\begin{align} -\delta \rho_{\mbox{\scriptsize{YM}}}[a^i] \delta
A^-, \label{ymrho}
\end{align} to be added to the induced charge from the source term
\begin{align} \delta \rho_{\mbox{\scriptsize{YM}}}+\delta
\rho^{(1)}+\delta\rho^{(2)}\equiv \delta \rho[a]. \label{99}
\end{align}
 As was observed in \cite{RGE}, $\delta \rho$ has a support in a
 narrow strip $1/b\Lambda^+ > x^- > 1/\Lambda^+$, while the
 original charges $\rho$ sit on
 $1/\Lambda^+ > x^- > 0$
  (Fig.~\ref{fig2}). Therefore,
  each step of quantum evolution piles up a layer of new
  classical charges at larger positive values of $x^-$.

The expansion of \begin{align}
S_{\mbox{\scriptsize{YM}}}[B^i+a^i+\delta A^i, a^-+\delta A^-]
\label{edm}
\end{align} requires a care because in the presence of quantum fluctuations
 the average field is not just $B^i$, but $B^i+
 \langle\delta A^i_{\mbox{\scriptsize{ind}}}[\delta \rho] \rangle$ \cite{RGE} where
  $\delta A^i_{\mbox{\scriptsize{ind}}}[\delta \rho]$ is that part
  of the soft field $\delta A^\mu$ induced by $\delta \rho$
   and obeys the Yang--Mills equation with the renormalized charge\footnote{In the LLA, only the transverse $\mu=i$
   components are important. Indeed, the one-- and the connected two--point functions
of $\delta A^i_{\mbox{\scriptsize{ind}}}$ are logarithmically
enhanced \cite{RGE} just like  the one-- and the two--point
functions of $\delta \rho[a]$. By expanding the left hand side of
Eq.~(\ref{newy}) to quadratic order in $\delta
A^i_{\mbox{\scriptsize{ind}}}$, we obtain relations between the
correlation functions:
 \begin{align} \sigma \equiv \langle \delta \rho[a] \rangle=\frac{\delta^2
 S_{\mbox{\scriptsize{YM}}}}{\delta A^-\delta A^i}\big\arrowvert_B
  \langle \delta A^{i}_{\mbox{\scriptsize{ind}}}\rangle+\frac{1}{2}
 \frac{\delta^3 S_{\mbox{\scriptsize{YM}}}}{\delta A^- \delta A^i \delta A^j }
 \big\arrowvert_B
 \langle \delta A^i_{\mbox{\scriptsize{ind}}} \delta A^j_{\mbox{\scriptsize{ind}}} \rangle \propto \alpha_s\delta
 \tau,
 \end{align} from the $\mu=+$ component of Eq.~(\ref{newy}) and  \begin{align} \frac{\delta^2 S_{\mbox{\scriptsize{YM}}}}{\delta
A^i\delta A^j}\big\arrowvert_B \langle \delta
A^{j}_{\mbox{\scriptsize{ind}}}\rangle+\frac{1}{2}
 \frac{\delta^3 S_{\mbox{\scriptsize{YM}}}}{\delta A^i \delta A^j \delta A^k }
 \big\arrowvert_B \langle \delta A^j_{\mbox{\scriptsize{ind}}} \delta
 A^k_{\mbox{\scriptsize{ind}}}\rangle =0, \label{ji}  \end{align}
 from the $\mu=i$ components. }
  \begin{align}
  \frac{\delta
 S_{\mbox{\scriptsize{YM}}}}{\delta A^\mu}\big\arrowvert_{B^i+\delta
 A^i_{\mbox{\scriptsize{ind}}}}= D_\nu F^{\nu\mu}\big\arrowvert_{B^i+\delta
   A^i_{\mbox{\scriptsize{ind}}}}=\delta^{\mu i}(\rho_{-\infty}+\delta
   \rho[a]). \label{newy}
   \end{align} For given $\delta \rho$, the solution of
   Eq.~(\ref{newy}) is again a pure gauge (see Eq.~(\ref{back}))
   \begin{align} B^i+\delta
   A^i_{\mbox{\scriptsize{ind}}}=\frac{i}{g}\bar{U}\partial^i\bar{U}^\dagger,
   \qquad \ \ \ \bar{U}^\dagger \equiv \mbox{P}\exp
\left(ig\int dx^- (\alpha + \delta \alpha)\right), \label{ubar}
\end{align} where $\delta \alpha$ is the induced field in the
Coulomb gauge.  Taking this into account,  we write
\begin{align} \delta A^i \to \delta A^i_{\mbox{\scriptsize{ind}}}[\delta \rho] + \delta
A^i,
\end{align}
  and expand Eq.~(\ref{edm}) around $B^i+\delta A^i_{\mbox{\scriptsize{ind}}}[\delta
  \rho]$. To linear order in small fields, we get
\begin{eqnarray} && S_{\mbox{\scriptsize{YM}}}[B^i+\delta A^i_{\mbox{\scriptsize{ind}}}+a^i+\delta A^i, a^-+\delta
A^-] \nonumber \\ &&\ \ =S_{\mbox{\scriptsize{YM}}}[B^i+\delta
A^i_{\mbox{\scriptsize{ind}}}]+D_{\mu}F^{\mu +}|_{B^i+\delta
A^i_{\mbox{\scriptsize{ind}}}}(a^-+\delta A^-) +D_{\mu}F^{\mu
i}|_{B^i+\delta A^i_{\mbox{\scriptsize{ind}}}}( a^i+\delta A^i)+
\cdots \nonumber \\&& \ \   =\left( \rho_{-\infty}+\delta \rho[a]
\right)
 (\delta A^- + a^-)+\cdots, \label{small}
 \end{eqnarray}
  where we used
 \begin{align}
  S_{\mbox{\scriptsize{YM}}}[B^i+\delta A^i_{\mbox{\scriptsize{ind}}}]=0, \end{align}
 because $B^i+ \delta A^i_{\mbox{\scriptsize{ind}}}$ is a pure
 gauge. Note that there is a mismatch between Eq.~(\ref{small})
 and Eq.~(\ref{replace}) (with the total induced charge Eq.~(\ref{99})).
  The term $\delta \rho[a] a^-$ in Eq.~(\ref{small}) is not cancelled
  by the source term. This term will play
an important role below.

 %For a later use,  we  isolate terms containing  $\delta A^{i,-}$
%fields
%\begin{eqnarray} &&S_{\mbox{\scriptsize{YM}}}[B^i+a^i+\delta A^i, a^-+\delta
%A^-] \nonumber \\ && \ \   \to S_{\mbox{\scriptsize{YM}}}[B^i+a^i,
%a^-] + (\rho+\delta \rho) \delta A^- + \delta \rho a^- -\delta
%\rho_{\mbox{\scriptsize{YM}}}\delta A^- +\cal{O}(\delta
%A^{i,-})^2. \label{ym}
%\end{eqnarray}

Collecting all factors, we are now prepared to integrate over the
semihard field. In the Gaussian approximation, Eq.~(\ref{22})
becomes
\begin{eqnarray} Z &=&\int
D\rho_{-\infty} W_\tau[\rho_{-\infty}]\int
 D\rho(x^+)D\delta A^{i,-} Da^{i,-} \nonumber \\
&& \qquad \times \exp\Bigl(iS_{\mbox{\scriptsize{YM}}}[B^i+\delta
A^i_{\mbox{\scriptsize{ind}}}[\delta \rho[a]]+\delta A^i,
 \delta A^-]+\frac{i}{2}
 a^\mu G^{-1}_{\mu \nu}a^\nu \nonumber \\
  && \qquad \qquad \qquad \qquad \qquad+i\delta \rho[a] a^- -i(\rho+\delta \rho[a])
  \delta A^- +iS_{\mbox{\scriptsize{WZ}}}[\rho] \Bigr), \label{44}
\end{eqnarray}
 where $G^{\mu\nu}$ is the background field propagator \begin{align} G^{-1}_{\mu\nu} \equiv \frac{\delta
 S_{\mbox{\scriptsize{YM}}}}{\delta A^\mu \delta A^\nu}
 \Big\arrowvert_{B^i}+\delta_{\mu -}\delta_{\nu -}\Pi.  \label{mod} \end{align}

We introduce the following trick.
\begin{eqnarray}
Z &=&\int D\rho_{-\infty}D\delta \rho_{-\infty}
W_\tau[\rho_{-\infty}]\int
 D\rho(x^+)D\delta \rho(x^+)D\delta A^{i,-} Da^{i,-} \ \delta (\delta \rho-\delta \rho[a]) \nonumber \\
&& \qquad \times \exp\Bigl(iS_{\mbox{\scriptsize{YM}}}[B^i+\delta
A^i_{\mbox{\scriptsize{ind}}}[\delta \rho]+\delta A^i,
 \delta A^-]+\frac{i}{2}
 a^\mu G^{-1}_{\mu \nu}a^\nu \nonumber \\
  && \qquad \qquad \qquad \qquad \qquad+i\delta \rho a^- -i(\rho+\delta \rho)\delta A^-
   +iS_{\mbox{\scriptsize{WZ}}}[\rho] \Bigr) \nonumber
   \\ &=& \int D\rho_{-\infty}D\delta \rho_{-\infty}
W_\tau[\rho_{-\infty}]\int
 D\rho(x^+)D\delta \rho(x^+)D\delta A^{i,-} Da^{i,-}D\pi
   \nonumber \\
&& \qquad \times \exp\Bigl(iS_{\mbox{\scriptsize{YM}}}[B^i+\delta
A^i_{\mbox{\scriptsize{ind}}}[\delta \rho]+\delta A^i,
 \delta A^-] +\frac{i}{2}
 a^\mu G^{-1}_{\mu \nu}a^\nu \nonumber \\ &&
   \qquad \qquad \qquad \qquad +i\delta \rho a^-
  +i\pi(\delta \rho-\delta \rho[a]) -i(\rho+\delta \rho)\delta A^-
   +iS_{\mbox{\scriptsize{WZ}}}[\rho]  \Bigr). \label{55}
\end{eqnarray}
 The integration over $a^{i,-}$ can be done. To the order of
interest (retaining terms enhanced by $\delta \tau$), we obtain
\cite{JKLW97}
\begin{eqnarray} Z&=&\int D\rho_{-\infty}D\delta \rho_{-\infty}
  W_\tau[\rho_{-\infty}]\int D\rho(x^+) D\delta \rho(x^+)
 D\delta A^{i,-} D\pi \nonumber \\&& \qquad \qquad \times \exp\Bigl(iS_{\mbox{\scriptsize{YM}}}[B^i+
 \delta A^i_{\mbox{\scriptsize{ind}}}[\delta\rho]+\delta A^i,
 \delta A^-] +i\pi(\delta\rho-\sigma)-\frac{1}{2} \pi \chi \pi
 \nonumber \\ && \qquad \qquad \qquad \qquad \qquad \qquad
  -i(\rho+\delta \rho) \delta A^- +iS_{\mbox{\scriptsize{WZ}}}[\rho]+
  iS_{\mbox{\scriptsize{WZ}}}[\delta \rho] \Bigr),
 \label{ii}
\end{eqnarray}
 where $\sigma=\langle \delta \rho[a]\rangle \propto \alpha_s\delta \tau$ and $\chi=\langle
 \delta \rho[a] \delta \rho[a]\rangle \propto \alpha_s\delta \tau$. Note that, in addition to the
 usual building blocks of the JIMWLK kernel, $\sigma$ and $\chi$,
 we have included a \emph{Wess--Zumino term for the induced charge}
$S_{\mbox{\scriptsize{WZ}}}[\delta
 \rho(x^+)]$.  This term does not contain a logarithm, but must be  included
 in order to render the theory at rapidity $\tau+\delta \tau$  gauge invariant.
 [The sum $-\delta \rho \delta A^-+S_{\mbox{\scriptsize{WZ}}}[\delta \rho]$
is gauge invariant.] We are not able to present a complete
derivation of $S_{\mbox{\scriptsize{WZ}}}[\delta \rho]$ here, but
argue how this term could arise. The origin of
$S_{\mbox{\scriptsize{WZ}}}[\delta
 \rho]$ is the elusive $\delta \rho a^-$ term. However,
 it  does \emph{not} arise in the Gaussian approximation. Indeed, if we performed a Gaussian integral in Eq.~(\ref{55}),
we would get an insignificant term
\begin{align}
 -\frac{i}{2} \int dx dy \delta \rho^a(x)G^{--}_{ab}(x,y)\delta \rho^b(y), \label{yoshi} \end{align}
  which is not the Wess--Zumino term, [Remember that, despite our notation, $S_{\mbox{\scriptsize{WZ}}}[\delta
 \rho]$ \emph{cannot} be written explicitly as a functional of $\delta \rho$ !] nor
 does it contain a logarithmic enhancement.\footnote{  Note that the $x^+$
  dependence
  of  $\delta \rho$ and $G^{--}$ originates from  that of the semihard field
  $a^\mu$. Therefore,
 the $p^-$ integral ($p_\perp^2/2b\Lambda^+ > p^- >
 p_\perp^2/2\Lambda^+$) in the Fourier representation of $G^{--}(x,y)$ contains an oscillating phase
  $e^{i(x^+-y^+)p^-}$ which prohibits  a
  logarithm.
  This is in contrast to Eq.~(\ref{cub}) where the integration over $y^+$ and $z^+$ can be
  explicitly performed  and a logarithm does arise.}  The absence of $S_{\mbox{\scriptsize{WZ}}}[\delta
 \rho]$  in the Gaussian approximation
  is understandable. As seen in Eq.~(\ref{wess}), the Wess--Zumino term is
  defined over the entire gauge group. Therefore, its origin must
have to do with
   gauge invariance. Once we truncate the Yang--Mills
    action in the Gaussian approximation, invariance under \emph{large} gauge
    transformations is lost. This suggests that in Eq.~(\ref{55})
    we have to perform an integral with the \emph{full} Yang--Mills
    action
 \begin{align} f[\delta \rho] \equiv \int Da^{i,-} \exp\bigl( iS_{\mbox{\scriptsize{YM}}}[a^i,a^-]+i\int
 dx^+ \delta \rho a^- \bigr), \label{kk} \end{align}
 where  we neglected the external fields $B^i$ and $\rho$. By this
 we formally treat the fields $a^{i,-}$ as large and assume that
 the external fields are not essential for the generation of $S_{\mbox{\scriptsize{WZ}}}[\delta
 \rho]$.
  [$S_{\mbox{\scriptsize{WZ}}}[\delta
 \rho]$ does not depend on $B^i$ or $\rho$.]
    Indeed, it is known \cite{diakonov2} that the Wess--Zumino term comes from the large $a^{i,-}$ region of the integral
    Eq.~(\ref{kk}).
 To see this, consider the following  change of integration
 variables in Eq.~(\ref{kk}) [See \cite{diakonov2} for more details.]
 \begin{align} a^{i,-} \to \tilde{a}^{i,-} =V^\dagger
 a^{i,-}V+\frac{i}{g}V^\dagger \partial^{i,-}V,  \label{large}
\end{align}
 where $V$ does not depend on  $x^-$.
 This is a gauge transformation which preserves the light--cone
 gauge condition $a^+=0$. Using the gauge invariance of the Yang--Mills action, one can easily
  check that
 \begin{align} f[\delta \rho]=f[V\delta \rho
 V^\dagger] e^{-\delta \rho V^\dagger \partial^- V}. \label{ext} \end{align}
  A general solution for Eq.~(\ref{ext}) is     \begin{align}
   f[\delta \rho]=g[ \delta
   \rho] e^{iS_{\mbox{\scriptsize{WZ}}}[\delta \rho]}, \label{eme} \end{align}
   where $g[\delta \rho]$
  is a gauge invariant function
   \begin{align}g[\delta \rho] =g[ V\delta \rho
   V^\dagger]. \end{align} In Eq.~(\ref{eme}), the Wess--Zumino term arises as a special
   solution to Eq.~(\ref{ext}). The relevance  of this term is easy to understand. The extra phase factor
   in Eq.~(\ref{ext}) due to
   the non-gauge invariance of $\int \delta\rho
a^-$ is precisely generated by the variation of the Wess--Zumino
term.  The function $g[\delta\rho]$ cannot be
   calculated exactly, but since the Gaussian approximation
   Eq.~(\ref{yoshi}) (with $G^{--}$ the free or dressed
   propagator)
  is already unimportant (i.e., not enhanced by a logarithm), we can simply set
   $g=1$ in the leading logarithmic approximation. This is the reasoning of our addition of the
   Wess--Zumino term in Eq.~(\ref{ii}).

 Returning to Eq.~(\ref{ii}), we observe that
\begin{align} S_{\mbox{\scriptsize{WZ}}}[\rho]+S_{\mbox{\scriptsize{WZ}}}[\delta \rho]
  = S_{\mbox{\scriptsize{WZ}}}[\rho + \delta \rho].  \label{wzreno}\end{align}
 This holds because $\rho$  and $\delta \rho$
  have different supports in
 $x^-$ \cite{RGE}: $1/\Lambda^+>x^->0$ for $\rho$, and $1/b\Lambda^+ > x^- > 1/\Lambda^+$ for $\delta \rho$.
Moreover, from Eq.~(\ref{ii}), we can easily deduce the evolution
of $W_\tau$ ($\rho'_{-\infty}\equiv \rho_{-\infty}+\delta
\rho_{-\infty}$)
\begin{align} W_{\tau+\delta
\tau}[\rho'_{-\infty}]=\int
D\pi\exp\left(i\pi(\delta\rho_{-\infty}-\sigma)
 -\frac{1}{2} \pi
 \chi \pi \right)W_\tau[\rho_{-\infty}],
 \label{fey}
 \end{align} in agreement with the path integral formula previously derived in
 \cite{rg}.\footnote{Here we neglect the $x^+$ dependence of
$\delta \rho$. See the remarks following Eq.~(\ref{wreal}).
Eq.~(\ref{fey}) shows that $\delta \rho_{-\infty}$ is Gaussian
distributed with the mean $\sigma$ and the  variance $\chi$
 (which are both static). }\footnote{ See, Eq.~(4.11) of
Ref.~\cite{rg}. There
 the authors work in the Coulomb gauge Eq.~(\ref{cou}). There is a
 subtlety in going from the light--cone gauge to the Coulomb gauge.
    The gauge function which realizes
 this rotation is not Eq.~(\ref{wil}), but Eq.~(\ref{ubar}). See the discussion in
Section 3 of the first paper in Ref.~\cite{RGE}.}
 The infinitesimal evolution Eq.~(\ref{fey}) is equivalent to the
 JIMWLK equation \begin{align} \frac{\partial}{\partial
 \tau}W_\tau[\rho]=\frac{1}{2}\frac{\delta}{\delta \rho_\tau}
 \frac{\delta}{\delta \rho_\tau}\big(\chi W_\tau[\rho]\big)-\frac{\delta}{\delta
 \rho_\tau}\big(\sigma W_\tau[\rho]\big), \label{JIMWLK} \end{align} where
 the subscript $\tau$ in $\rho$  means that the derivatives are
 taken at the highest value of $x^-$; $x^-=1/\Lambda^+$ \cite{RGE}.
 [Remember that the support of $\delta \rho_{-\infty}$ in Eq.~(\ref{fey}) is
  $1/b\Lambda^+ > x^+> 1/\Lambda^+$.]
 Using Eq.~(\ref{wzreno}) and Eq.~(\ref{fey}),   we finally arrive at
  \begin{eqnarray} Z&=&\int
 D\rho_{-\infty}' W_{\tau+\delta
 \tau}[\rho_{-\infty}']  \int D\rho'(x^+)\int_{\tau + \delta \tau} D\delta A^{i,-}
  \nonumber \\&& \qquad \qquad \times \exp\left(iS_{\mbox{\scriptsize{YM}}}
 [B'^i+\delta A^i,
 \delta A^-] -i\rho'\delta A^- +iS_{\mbox{\scriptsize{WZ}}}[\rho'(x^+)]
  \right), \label{rgg} \end{eqnarray} where
   $B'^i=B^i+\delta  A^i_{\mbox{\scriptsize{ind}}}[\delta \rho]$.
   Eq.~(\ref{rgg}) is exactly the same form as our starting
  formula Eq.~(\ref{fin}). This is the JIMWLK  renormalization group picture. Namely,
   the quantum effects at
  one step of evolution is completely absorbed by the change of
  the weight function and the form of the effective theory is
  preserved.

\section{Conclusions  }
\setcounter{equation}{0}
 In this paper, we have investigated the role of the Wess--Zumino
 term in the high energy limit of QCD where the target becomes a Color Glass Condensate.
  We have shown that the Wess--Zumino term can be naturally
  incorporated in the JIMWLK formalism as a part of the source term
  $S_{\mbox{\scriptsize{W}}}$. The simple eikonal coupling between the gauge field and
   the source $-\rho\delta
  A^-$ makes the renormalization group description more
  transparent.  We have argued that after
  one step of quantum evolution the functional integral Eq.~(\ref{fin}) remains
  the same form  including the source term. This last point has
  not been discussed in
   the existing proofs of renormalization
  group \cite{JKLW97,RGE}.

  Finally, we expect that the real
  strength of our approach is that it can be straightforwardly applied to other limits of
  high energy QCD, namely, the dilute regime (Fig.~1(b)) and
  the Pomeron loop regime (Fig.~1(c)), where the importance of the Wess--Zumino term
  has been
  first recognized.
 The color glass averaging in the dilute regime proposed in
 \cite{KL1} is (see also \cite{KL5})
\begin{align} \langle \hat{\rho}^a \hat{\rho}^b \hat{\rho}^c\cdots
\rangle_\tau =\int D\rho(t) W_\tau[\rho(t)]
\rho^a(t_1)\rho^b(t_2)\rho^c(t_3) \cdots, \label{ave}
 \end{align} where the ordering variables $t_i$  are constrained such that
 $ t_1 > t_2 > t_3 > \cdots$, but otherwise
 arbitrary. The ``weight function" $W_\tau[\rho(t)]$ contains a
 complex phase
 $e^{iS_{\mbox{\scriptsize{WZ}}}[\rho(t)]}$ which ensures the
 color commutators of
  $\hat{\rho}$.
  It follows from  our analysis in Section 2 (see Eq.~(\ref{start})) that Eq.~(\ref{ave}) should read \begin{align} \langle \hat{\rho}^a \hat{\rho}^b \hat{\rho}^c\cdots
\rangle_\tau =\int D\rho_{\pm \infty}W_{\tau}[\rho_{ \pm
\infty}]\int_{\rho_{-\infty}}^{\rho_{\infty}}
D\rho(x^+)e^{iS_{\mbox{\scriptsize{WZ}}}[\rho(x^+)]}
\rho^a(x^+_1)\rho^b(x^+_2)\rho^c(x^+_3) \cdots, \label{yoshi}
 \end{align} with the \emph{same} Wess--Zumino term appearing in
 Eq.~(\ref{sou}).\footnote{Note that the approximation
 Eq.~(\ref{wreal}) is invalid in the dilute regime. This means
 that $W_\tau[\rho_{\pm \infty}]$ is complex even without the Wess--Zumino term.}
  Therefore, the JIMWLK formalism
 with our new source term allows us to treat different limits of
 high energy QCD in a single framework.
    The renormalization
 group description of quantum evolution in the dilute regime is
 subtler
 than in the JIMWLK regime because the division between the ``quantum" and ``classical" theories
 is shifted towards the quantum side. [Compare Eq.~(\ref{yoshi}) with Eq.~(\ref{observable}).] We leave this
 problem for future works.

\section*{Acknowledgments}
%\vspace*{-0.5cm}
I would like to thank
 Edmond Iancu, Larry McLerran, Anna Sta\'sto
 and Dionysis Triantafyllopoulos for  discussions and the previous collaboration
 that paved the way for the present results.
  I am especially grateful to Edmond Iancu and Larry
McLerran for valuable help and encouragement. I also thank Jochen
Bartels, Kenji Fukushima and Alex Kovner for discussions. This
research has been supported by RIKEN, Brookhaven National
Laboratory and the U.~S.~Department of Energy  [Contract No.
DE-AC02-98CH10886].

\appendix

 \section{Remarks on the saddle point solution Eq.~(\ref{saddle}) } \label{A2}
 For possible future applications, in this appendix we note  a curious property of the saddle point
   solution Eq.~(\ref{saddle}). First we observe that the source
   term vanishes at the saddle point \begin{align} S_{\mbox{\scriptsize{W}}}
   =-\int dx^+ \rho(x^+) \delta A^-+
   S_{\mbox{\scriptsize{WZ}}}[\rho(x^+)]\Big\arrowvert_{\mbox{\scriptsize{saddle
   point}}}=0. \label{00} \end{align} To show this, return to Eq.~(\ref{rho1})
   \begin{align}
\rho^a(x^+)=-\frac{gJ}{2}\Tr \tau^3S \tau^a S^\dagger
=(W\rho_{-\infty}W^{\dagger})^a, \qquad
\rho^a_{-\infty}=-\frac{gJ}{2}\Tr \tau^3S_{-\infty} \tau^a
S^\dagger_{-\infty}.
 \end{align} We see that $S(x^+)$ and $S_{-\infty}$ are related by
 \begin{align} S(x^+)=S_{-\infty}W^\dagger(x^+). \end{align}
 Therefore, the Wess--Zumino term Eq.~(\ref{wess}) becomes
\begin{eqnarray}
 &&S_{\mbox{\scriptsize{WZ}}}[\rho(x^+)]\big\arrowvert_{\mbox{\scriptsize{saddle point}}}=iJ\int dx^+ \tr \,
 [\tau^3 S_{-\infty} W^\dagger \partial^-WS^\dagger_{-\infty}]
 \nonumber \\ && \qquad \qquad \qquad=-gJ\int dx^+ \tr \,
 [\tau^3 S_{-\infty} W^\dagger
 \tau^aWS^\dagger_{-\infty}]\frac{\delta A^-_a}{2}=\int dx^+
 \rho^a(x^+) \delta A^-_a.
 \end{eqnarray} Since $S_{\mbox{\scriptsize{W}}}$ is gauge
 invariant, Eq.~(\ref{00}) holds in any gauge. Then let us consider the exponential factors in
 Eq.~(\ref{fin})
  in
 the \emph{Coulomb gauge} Eq.~(\ref{cou}) \begin{align}
 S_{\mbox{\scriptsize{YM}}}[\alpha,\delta \tilde{A}^-]-\int \tilde{\rho}\delta \tilde{A}^- +
  S_{\mbox{\scriptsize{W}}}[\tilde{\rho}(x^+)], \label{tree} \end{align}
 where we neglected $\delta A^i$.
   In the context of \cite{HIMST},  Eq.~(\ref{tree}) should be called the \emph{tree level effective
 action}. Since the only non-vanishing  components of the field strength
 are $F^{\pm i}$, the Yang--Mills action becomes \begin{align}
 -\frac{1}{4}\int \tilde{F}^{\mu \nu}\tilde{F}_{\mu \nu} \approx
 \int \tilde{F}^{+i}\tilde{F}^{-i} =\int \partial^i \alpha
 \partial^i \delta \tilde{A}^- =-\int (\nabla^2 \alpha) \delta  \tilde{A}^-=\int
 \tilde{\rho}
 \tilde{A}^-. \end{align} Therefore, in the Coulomb gauge, at the saddle point Eq.~(\ref{saddle}),
  all the
 three terms in Eq.~(\ref{tree}) are equal up to a sign. In
 particular, \begin{align}
 S_{\mbox{\scriptsize{YM}}}=S_{\mbox{\scriptsize{WZ}}}.  \label{conjecture} \end{align}
 Eq.~(\ref{conjecture}) was conjectured in \cite{KL1} as a possible origin of the Wess--Zumino
 term. Although at the moment we
 do not see a connection between our derivation and the argument in
 \cite{KL1}, in any case it is interesting to pursue physical
 implications of Eq.~(\ref{conjecture}).

\end{document}